\documentclass[manuscript,smallcondensed,smallextended]{svjour2}
\usepackage{graphicx}
\usepackage{color}

\begin{document}

\title{Nonlocal appearance of a macroscopic angular momentum}

\author{G.~S.~Paraoanu \and R. Healey}
\institute{G. S. Paraoanu \email{paraoanu@cc.hut.fi}
\at Low Temperature Laboratory, Aalto University, P. O. Box 15100, FI-00076
AALTO, Finland
\at
Institute for Quantum Optics and Quantum Information (IQOQI),
Austrian Academy of Sciences, Boltzmanngasse 3, A-1090 Vienna, Austria
\and
R. Healey \email{rhealey@email.arizona.edu}
\at
Department of Philosophy, University of Arizona, Tucson AZ 85721, U.S.A.
\at
Institute for Quantum Optics and Quantum Information (IQOQI),
Austrian Academy of Sciences, Boltzmanngasse 3, A-1090 Vienna, Austria}

\maketitle

\begin{abstract}
We discuss a type of measurement in which a macroscopically large angular
momentum (spin) is ``created'' nonlocally by the measurement of just a few
atoms from a double Fock state. This procedure apparently leads to a blatant
nonconservation of a macroscopic variable - the local angular momentum.
We argue that while this {\it gedankenexperiment} provides
a striking illustration of
several counter-intuitive features of quantum mechanics, it does not imply a
non-local violation of the conservation of angular momentum.
\end{abstract}


\section{Motivation}

Lalo\"{e} \cite{laloe} has recently pointed out a beautiful paradox in
quantum physics. Suppose we have two Bose-Einstein condensates (BEC)  polarized in opposite
directions along the $z$-direction \cite{ketterle} and we perform a
measurement of the
spin (angular momentum) along a transversal direction by absorbing particles
emerging from one cloud or the other.  Then it can be shown that a macroscopic angular momentum along a
random transversal direction appears after the detection of only a few particles \cite{theory}. Moreover, owing to the fact that the
clouds can be imagined as having a very large spatial extension, apparently
a large macroscopic angular momentum can be made to appear almost
instantaneously at a distance. The effect, as in the Hanbury-Brown Twiss
experiment, may be attributed to the lack of which-path
information about the particles (we cannot know from which cloud they
emerged). In this paper we show that this situation is by no means singular  in
quantum physics. The paradox originates from our tendency to attribute
reality to the eigenvalue/eigenvector of an observable whenever we are in a
position to predict with certainty the result of measuring it.
The  paper is organized as follows. In Section \ref{cons} we give a general discussion of conservation laws in quantum mechanics, with emphasis on the role of measurement.
In Section \ref{first} we give a brief review of the \textit{gedankenexperiment} of Lalo\"{e}. The resolution of the non-conservation paradox is presented in
Section \ref{res}. Finally, in Section \ref{conc} we discuss the implications of the above for the issue of ``reality'' in quantum physics.

\section{Measurement and conservation laws in quantum mechanics}
\label{cons}

In this section we attempt to clarify  what is meant by conservation laws in quantum mechanics, and in particular how we are to interpret these laws as measurement enters the picture.

Both in quantum and classical mechanics, conserved
quantities are associated with dynamical symmetries. In classical mechanics,
a dynamical variable $A$ of a system ${\cal S}$ always has some precise
real-numbered value $a$, and $A$ is conserved in any process involving ${\cal S}$
just as long as the value of $A$ continues to equal $a$. Suppose that $A$
has no explicit time-dependence and that ${\cal S}$ is subject to a conservative
process governed by a Hamiltonian $H$ with no explicit time-dependence. Then
$A$ is a conserved quantity of ${\cal S}$ if and only if $\left\{ A,H\right\} =0$, where $\left\{ \cdot , \cdot \right\}$ is the Poisson bracket.

In quantum mechanics, a dynamical variable $A$ of a system ${\cal S}$ is not
assumed always to have some precise real-numbered value. Nevertheless, if $A$
is represented by a self-adjoint operator $\hat{A}$ with no explicit
time-dependence and ${\cal S}$ is subject to a conservative process governed by a
Hamiltonian $\hat{H}$ with no explicit time-dependence, then $A$ is said to
be conserved if and only if it commutes with the Hamiltonian, $\left[ \hat{A},\hat{H}\right] =0$. In
the Schr\"{o}dinger picture ${\cal S}$'s pure quantum state satisfies $\left\vert
\psi (t)\right\rangle =\hat{U}(t)\left\vert \psi (0)\right\rangle $, where $\hat{U}(t)=\exp \left[ - i\hat{H}t/\hbar\right]$.
Then for the expectation value we have
\begin{eqnarray}
\left\langle A\right\rangle _{t}&\equiv&
\left\langle \psi (t)\left\vert \hat{A}\right\vert \psi (t)\right\rangle
=\left\langle \psi (0)\hat{U}^{\dag }(t)\left\vert \hat{A}\right\vert \hat{U}(t)
\psi (0)\right\rangle \label{av1}\\
&=&\left\langle \psi (0)\left\vert \hat{A}\right\vert
\psi (0)\right\rangle =\left\langle A\right\rangle _{0},\label{av2}
\end{eqnarray}
provided that
$\left[ \hat{A},\hat{H}\right] =0$.
The same result follows immediately in the Heisenberg picture, where we can write
$\hat{A}(t) \equiv \hat{U}^{\dag }(t)\hat{A}\hat{U}(t) = \hat{A}(0)$. Similarly, all higher moments of $A$ are
conserved while the state of $\cal{S}$ evolves unitarily according to $\hat{U}
(t)$. If one now uses the spectral theorem to expand
$\hat{A}$ in eigenvalues,
\begin{equation}
\hat{A}=\sum_{j}a_{j}|a\rangle _{jj}\langle a|,
\end{equation}
 one notices immediately that  the probability distributions corresponding to any measurement result $a_{j}$ are conserved,
\begin{equation}
P_{j,|\psi\rangle}(t) = |_{j}\langle a|\psi (t)\rangle|^{2} = P_{j,|\psi\rangle}(0),
\end{equation}
 owing to the fact that the $|a\rangle_{j}$'s are also eigenvectors of the Hamiltonian $\hat{H}$. This provides another way to derive Eq. (\ref{av1}-\ref{av2}), namely $\left\langle A\right\rangle _{t} = \sum_{j} a_{j} P_{j}(t) = \sum_{j} a_{j} P_{j}(0) =
\left\langle A\right\rangle _{0}$.

The definitions above only allow us to predict that the probability distribution of a conserved observable will not change after a unitary (Hamiltonian) evolution. They do not say anything about nonunitary processes, in particular about
what we are allowed to say when a {\it single measurement} of a conserved quantity is performed.
Since the work of Dirac and Von Neumann it has been common to regard
measurement in quantum mechanics as a \textit{non}unitary process that
projects a system's pure quantum state onto an eigenstate $\left\vert
a\right\rangle $ with eigenvalue $a$ equal to the result of the measurement. These processes are called projective measurements.
Let us consider in more detail the following situation. Suppose that at $t=0$ we measure $A$ and we find the value $a_{k}$. Then at $t=0$
we can assign the state $|a\rangle _{k}$. We let the system evolve with
the evolution operator $\hat{U}(t)=\exp [-i\hat{H}t/\hbar]$ into $\hat{U}
(t)|a\rangle _{k}$. Then, at a time $t$, the probability of obtaining the
result $a_{j}$ is
\begin{equation}
P_{j,k}(t)=|_{j}\langle a|\hat{U}(t)|a\rangle _{k}|^{2}
\end{equation}%
This probability is $P_{j,k}(t)=\delta _{jk}$ only if $\left[ \hat{A},\hat{H}%
\right] =0$. This means that only in this case will we find (with certainty)
the same value for the variable when we measure it again at the time $t$.
This makes us think that $A$ has value $a_{k}$ in state $\hat{U}(t)|a\rangle
_{k}$: That the variable is conserved not only in the sense that if we
measure it we certainly find this same value, but that the value in some
sense exists there between the measurements.

We can make this thought more precise by introducing the following interpretative principle:

\vspace{2mm}
{\bf Descriptive Completeness:}

\noindent{\it A dynamical variable $A$ of a system ${\cal S}$ has precise real-numbered value $%
a_{k}$ if and only if the pure quantum state of $A$ is an eigenstate $\left\vert
a\right\rangle_{k}$ with eigenvalue $a_{k}$.}
\vspace{2mm}

The statement above makes clear that there exists a distinction between the state of a system being an eigenstate of an operator (observable) and that observable actually assuming the corresponding eigenvalue. The Descriptive Completeness principle  associates the
two: it is a rather minimalistic statement of completeness of quantum theory, by proposing a one-to-one correspondence between only certain mathematical entities that appear in the theory and
ontological entities. As we will show here, this interpretative principle, although it looks innocuous and natural, is at the core of the paradoxes we discuss. Thus quantum mechanics is not a descriptively complete theory, in the sense defined above.

Let us now examine some of the consequences of this rule, when used in conjunction with the quantum-mechanical unitary and nonunitary processes.

{\it Unitary evolution:} If this descriptive completeness rule is a valid way to interpret unitary evolution in quantum mechanics, the immediate consequence when analyzing conserved variables can be formulated as follows:
If $A$ has
value $a$ at
$t_{1}$ on ${\cal S}$ and is conserved throughout the interval between $t_{1}$and $%
t_{2}$, then $A$ has value $a$ at $t_{2}$; while if $A$ has \textit{%
no }value $a$ at $t_{1}$ and is conserved through the interval
between $t_{1}$ and $t_{2}$, then $A$ has \textit{no }value $a$ at $t_{2}$ - unitary evolution neither creates nor alters the value of a conserved
variable.

{\it Nonunitary processes:} For nonunitary processes, a conservation law for dynamical variables can be formulated simply as: a
dynamical variable $A$ is conserved in a projective measurement process on $\cal{S}$ if and only if, when applied to an initial eigenstate of $\hat{A}$ with
eigenvalue $a$, the process results in a final eigenstate of $\cal{S}$ also with
eigenvalue $a$. If this is used in conjunction with the Descriptive Completeness rule, we end up with the statement that if $A$ has value $a$ at $t_{1}$ on
${\cal S}$ and is
conserved in a projective measurement from $t_{1}$to $t_{2}$, then $A$ has
value $a$ at $t_{2}$. But, unlike for unitary evolution, if $A$ has \textit{no} value $a$ at $t_{1}$ and is conserved in a projective measurement between $t_{1}$ and $t_{2}$,
then $A$ \textit{does} have some\textit{\ }value $a$ at $t_{2}$: in
this case, a dynamical variable acquires a value on projective measurement
\textit{even if that variable is conserved during the process}. A
dynamical variable then takes on a precise real-numbered value as a result
of a projective measurement in which that variable is conserved.

Note that nowhere here we have said anything about where in space the measurement happens. The claim of existence of a value is made for the entire system under discussion, irrespective of how large its spatial extension is and of the
fact that the measurement process itself may be localized to only a small part of the system. The consequence of this is nonlocality: apparently, if the Descriptive Completeness rule were to hold, a projective measurement on a subsystem can produce instantaneously a value for a variable of another subsystem, no matter how far away. Clearly, this would violate conservation laws locally.

Lalo\"{e}'s \textit{gedankenexperiment} provides a dramatic illustration of
three peculiar consequences of the implicit use of Descriptive Completeness for the
understanding of measurement and
conservation in a particular setup: A distant measurement on a double condensate may prompt the
\textit{nonlocal} appearance of some definite value of a conserved dynamical
variable on a system, this value may be uniquely determined by the
post-measurement state of that system, and (most striking) this value may be
macroscopic, even though a distant microscopic measurement prompted its
nonlocal appearance. Aspects of the situation described by Lalo\"{e} are far from being unusual
in the realm of quantum mechanics. The first two consequences should be familiar
from earlier work in the foundations of quantum physics related to the Einstein-Podolsky-Rosen (EPR) paradox, to Bell states, and to Greenberger-Horne-Zeilinger (GHZ) states \cite{EPR,Bohm,GHZ}.
However, the third consequence, the apparent nonlocal emergence of a macroscopic value as a consequence of a distant microscopic measurement process, is new.

\section{Lalo\"e's \textit{gedankenexperiment}}
\label{first}

\begin{figure}[t]
\begin{center}
\includegraphics[width=9cm]{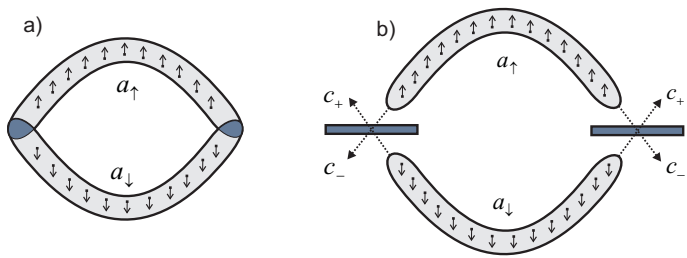}
\end{center}
\caption{(a) Two condensates prepared in a double Fock spin state, and overlapping in two regions of space. Photon absorption measurement in one region results in the establishment of a macroscopic spin in a direction perpendicular to the initial polarization. (b) Equivalent measuring system using beam splitters.}\label{figure1}
\end{figure}

Let us now briefly review the situation analyzed in \cite{laloe}. Consider
BEC gases with atoms in two internal states, see Fig. (\ref{figure1}). Because any two-level system is mathematically equivalent to a
spin-1/2, we can regard the two states as being the states "up" $|\uparrow \rangle $ and "down" $|\downarrow\rangle $ of a spin along the $z$-direction.  We will denote from now on by $|\pm
\rangle _{\vec{n}}$ the two eigenvectors of the spin-1/2 along a direction $\vec{n}$, in other words $(\vec{n}\cdot\vec{\hat{\sigma}})|\pm \rangle _{\vec{n}}=\pm
|\pm \rangle _{\vec{n}}$. In particular, $|\uparrow \rangle =|+\rangle _{z}$
and $|\downarrow \rangle =|-\rangle _{z}$. Now  we prepare independently two
condensates, one in which all the particles are in the state $|\uparrow
\rangle $, and the other one in which they are in the state $|\downarrow
\rangle $. For simplicity we take the number of particles in each state as
equal $N_{\uparrow }=N_{\downarrow }=N/2$, where $N$ is the total number of
particles. The system can then be described by a double Fock state
\begin{eqnarray}
|\Psi \rangle _{\mathrm{initial}}&=&|N_{\uparrow }\rangle |N_{\downarrow
}\rangle \\ &=&\frac{1}{\sqrt{N_{\uparrow }!}\sqrt{N_{\downarrow }!}}\left[\hat{a}_{\uparrow
}^{\dag}\right]^{N_{\uparrow }}
\left[\hat{a}_{\downarrow }^{\dag}\right]^{N_{\downarrow }}|\mathrm{vac} \label{ini}
\rangle ,
\end{eqnarray}
where $\hat{a}_{\downarrow}$, $\hat{a}_{\downarrow }^{\dag}$ and $\hat{a}_{\uparrow}$, $\hat{a}_{\uparrow}^{\dag}$ are the respective bosonic annihilation and creation operators
for spin-down and spin-up atoms.
Now the atoms are detected such that which-path information
cannot be obtained (\textit{i.e.} it is not possible to identify from which
cloud, $\uparrow $ or $\downarrow $, the particles came). In the real interference experiment \cite{ketterle}, this was achieved by absorption of photons in a certain region in space where the condensates overlap (Fig. \ref{figure1}a). The effect may be easier to understand if one thinks in terms of a two-channel beam-splitter measurement (Fig. \ref{figure1}b). In this case, the
measurement of a single particle can be described by the action of the
operators
\begin{equation}
\hat{c}_{\pm }=\frac{1}{\sqrt{2}}(\hat{a}_{\uparrow }\pm \hat{a}_{\downarrow }).
\end{equation}
This is in fact a measurement of the spin along the $x$
direction of particles coming from the two clouds. The signs $\pm $
correspond to finding a particle with spin-component $\pm $ along the $x$
axis. To convince ourselves that this is the case, we recall that in
second quantization the number of particles operator is
\begin{equation}
\hat{n}=\hat{a}_{\uparrow }^{\dag }\hat{a}_{\uparrow }+\hat{a}_{\downarrow }^{\dag }\hat{a}_{\downarrow },
\end{equation}%
and the spin along the $x$ direction is
\begin{equation}
\hat{\sigma} _{x}=\hat{a}_{\uparrow }^{\dag }\hat{a}_{\downarrow }+\hat{a}_{\downarrow }^{\dag
}\hat{a}_{\uparrow }.
\end{equation}
We can immediately prove the identity
\begin{equation}
\frac{1}{2}(\hat{n}\pm \hat{\sigma} _{x})=\hat{c}_{\pm }^{\dag }\hat{c}_{\pm }. \label{smur}
\end{equation}%
Since in our experiment the particles are  detected one at a time, we recognize
on the left-hand side of Eq. (\ref{smur}) the first-quantized projection operator
$\frac{1}{2}(I\pm \hat{\sigma} _{x})$ for the measurement of spin in the $x$ direction with results $\pm$. The left-hand side is the number operator for detection in the $\pm$ channels in Fig. (\ref{figure1}b). Thus, the measurement along the $x$ direction is realized by counting the detections in each of these channels.
The surprising result  is that the repeated application of  $\hat{c}_{\pm }$ (or $\hat{c}_{\pm }^{\dag }\hat{c}_{\pm }$ if one wants to
conserve the number of particles) results in the appearance of
phase coherence between the $\uparrow $ and $\downarrow $ components.
The number of detection events $N_{d}$ after which this phase is established does not have to be macroscopic for the system to acquire phase-coherence; typically after only a few tens of  detections the system approaches to a good approximation the state
\begin{equation}
|\Psi \rangle _{\mathrm{after}}=\frac{1}{\sqrt{(N-N_{d})!}}\hat{c}_{\varphi
}^{\dag (N-N_{d})}|\mathrm{vac}\rangle.  \label{after}
\end{equation}%
Here
\begin{equation}
\hat{c}_{\varphi }=\frac{1}{\sqrt{2}}(\hat{a}_{\uparrow }+e^{i\varphi }\hat{a}_{\downarrow })
\end{equation}%
is the annihilation operator along a direction given by the unit vector $\vec{n}_{\varphi}$ in the $xOy$ plane which
makes some angle $\varphi $ with $Ox$. The state Eq. (\ref{after}) is an
eigenvalue of the total spin component $(\hbar/2)\sum_{i=1,N-N_{d}}\vec{\hat{\sigma}}\cdot\vec{n}_{\varphi}$
along this direction, with a macroscopic value $(\hbar /2)(N-N_{d})$.
Indeed, $(\hbar /2)\hat{c}_{\varphi }^{\dag }\hat{c}_{\varphi }|\Psi \rangle _{\mathrm{
after}}=(\hbar /2)(N-N_{d})|\Psi \rangle _{\mathrm{after}}$.
The phase $\varphi$ emerging after this sequence of measurements is random, in the sense that repeating the experiment produces in general a different value for $\varphi$. Note that the initial state Eq. (\ref{ini}) does not have a well-defined phase along any direction. Here we have a situation in  which apparently a microscopic process (note that $N_{d}$ can be much smaller than $N$) led to the appearance of a macroscopic quantity.

Moreover, the two clouds may be imagined to overlap
only in two widely separated regions, as in Fig. (\ref{figure1}). In this case,
the $N_{d}$ detections in one region prompts the almost
instantaneous appearance at a distance of a macroscopic angular momentum in the other region.

\section{Nonlocal creation of angular momentum?}
\label{res}

Let us now discuss in more detail the mechanism by which this macroscopic
angular momentum is apparently created nonlocally. We show that in the end this
paradox is not conceptually different from the situation of standard EPR-type experiments.


As before, the measurement process is described only by the operator $%
\hat{c}_{\pm }$. Before the first detection the probabilities to obtain the
results $\pm $ are equal, 1/2. Suppose now that the first particle is
detected and the result is $+$. Then, after applying the operator $\hat{c}_{+}$ to
$|N_{\uparrow }N_{\downarrow }\rangle $, we get the many-body wavefunction
\begin{equation}
\frac{1}{\sqrt{N}}\left[ \sqrt{N_{\uparrow }}|N_{\uparrow }-1,N_{\downarrow
}\rangle +\sqrt{N_{\downarrow }}|N_{\uparrow },N_{\downarrow }-1\rangle %
\right]   \label{wf}
\end{equation}
We now ask again what is the probability of getting the results $\pm $;
these can be calculated by applying $c_{\pm }$ on the wavefunction Eq. (\ref%
{wf}) and taking the square modulus of each of the two results. In the limit
of large number of particles, $N\gg 1$, we find that the probability of
obtaining the result $+$ is now 3/4 while that of obtaining the result $-$
is 1/4. Thus the previous detection, with the result $+$, has modified the
probabilities for the next detection, favoring a $+$ result. Now if a $+$
result is obtained again, the probability to find $+$ at the next detection is
5/6. In general, the probability to obtain $+$ after previous $k$
consecutive $+$ detections is $(2k+1)/2(k+1)$, which very quickly approaches
1. Thus after only a few consecutive $+$ detections, the wavefunction
becomes very close to an eigenvector of the operator $c_{+}$.

Let us now do the same calculation in the situation where we know where the
particles come from. For example, we either block one of the condensates,
move it further, or controllably release particles from either one of them.
Suppose that for the detection of the first particles we choose the condensate
with $\uparrow $, and suppose, as before, that we get the result $+$. We
again find the state of the system after this detection by applying $c_{+}$,
but this time on $|N_{\uparrow }\rangle |\mathrm{vac}\rangle $. For the next
detection, we can choose to use either $\uparrow $ atoms or $\downarrow $
atoms. In both cases, we obtain equal probabilities for the results $\pm $!
In general, after $k_{\uparrow }$ atoms have been extracted from the cloud
$\uparrow $ and $k_{\downarrow }$ from the cloud $\downarrow $, the overall
state of the system remains a Fock state $|N_{\uparrow }-k_{\uparrow
},N_{\downarrow }-k_{\downarrow }\rangle $.

Thus the essential ingredient for obtaining the phase
is lack of {\it which-path} information about which cloud
the detected particles have originated from.
This is embedded in the definition of the operators $\hat{c}_{\pm}$ as superpositions
between $\hat{a}_{\uparrow}$ and $\hat{a}_{\downarrow}$. It almost
looks like our ignorance  has created a macroscopic spin
component in the $xOy$ plane! Of course, the two situations (availability or not of which-way information) correspond to physically different setups. But these setups are local, and, if we accept the Descriptive Completeness principle, it seems possible to determine, from a spatially separated location, whether a macroscopic element of reality pops up into existence or not.

Our resolution of the ``paradox'' appeals to two separate measurement setups. The first one is Alice's, who measures transverse components of spin on a relatively small number of particles in a tiny region where the two clouds overlap: While in the second, Bob measures the total transverse spin-component of the overlapping clouds in a large, distant region along just the direction in which this is then almost certain to record a macroscopic result. If we treat Alice's measurement as projective, then it puts the total condensate into a state which is very close to an eigenstate of spin-component along the direction Bob happens to measure, corresponding to a macroscopic eigenvalue. Now, when Bob measures this spin-component on the large cloud of condensate in his region, far from Alice, he is almost certain to get a macroscopic result. But this does not mean that Alice's measurement produces a  ``real'' macroscopic value of angular momentum in Bob's region. Bob's measurement does not simply reveal this macroscopic value of a pre-existing spin-component of or in the cloud produced by Alice's projective measurement. Instead, the macroscopic spin-component in Bob's region ``emerges'' during Bob's measurement following an interaction with Bob's measuring device. There is no reason to suspect that this local interaction involves any violation of the conservation of angular momentum. Certainly, nothing that happens near Alice creates a macroscopic angular momentum near Bob in violation of local  conservation of angular momentum. Thus our conclusion is that Lalo\"{e}'s gedanken-experiment involves no nonlocal failure of the conservation law
of angular momentum, but rather that it reveals that Descriptive Completeness is not a valid principle in quantum physics.



To revisit the structure of the argument,  we emphasize the following two
key points:

1. in the natural way of extending  the standard analysis of conservation of an
observable to cover the case of projective measurement, a conserved observable with no
prior definite value may acquire an eigenvalue on projective measurement,

2. but after a subsystem ${\cal S}_{1}$ of a composite system ${\cal S}_{1}\otimes{\cal S}_{2}$ is put into an eigenstate of an observable
following a projective measurement on a distinct subsystem ${\cal S}_{2}$, one cannot immediately
infer that the observable has the corresponding eigenvalue on ${\cal S}_{1}$: one can infer only that a
measurement of that observable will certainly reveal that value. So an inference from a quantum state
to a value of an observable is not always valid even when that state is an eigenstate of
that observable.

\section{Conclusions and further discussions: does angular momentum exist?}
\label{conc}

Franck Lalo\"{e} \cite{laloe} notices the apparent nonconservation of
angular momentum for a double Fock state and concludes that (barring
controversial interpretations such us Everett's) we are almost
inevitably forced to agree with the EPR conclusion that this state does not provide a complete description of the physical system ({\it i.e.} Descriptive Completeness fails).

If the wavefunction is not an element of reality (or a representation of
one) but just a tool to predict the results of measurements,
then there is no paradox. The macroscopic angular momentum simply appears in
the very act of measuring a macroscopic angular momentum. Note that this is
true \textit{whether or not} the measured system's pre-measurement quantum
state is an eigenstate of that angular momentum (or other\ quantity) that is
measured. In the Stern-Gerlach apparatus the angular momentum does not
appear until the particles interact one by one with the detector after
passing the magnets. In this example, the initial wavefunction is not an
eigenstate of spin component ($\pm 1$) along $x$ but a superposition. On the
other hand, in Lalo\"{e}'s example the wavefunction after the distant
microscopic measurements but before a local measurement of angular momentum
is (very nearly) an eigenstate of angular momentum along a direction in the $%
xOy$ plane which makes some angle $\varphi $ with $Ox$. In this respect the
situation described by Lalo\"{e} mirrors the GHZ states: there, following
(two) distant measurements we also have an eigenvalue of a local observable
to which we cannot attribute physical reality prior to its measurement
 even
when the state of the system is an eigenstate of that observable.

This of course leaves us with the question: When a system is prepared in an
eigenstate of an observable, what ``exists'' out there? The only answer that standard quantum
theory provides
\cite{peresbohr}  is simply: This wavefunction refers to an ensemble of identically
prepared systems, on which a device for measuring that observable will
record the (same) corresponding eigenvalue for each system. This in no way
implies that this value \emph{exists} as an intrinsic property of each of
the components of the ensemble, an implication that also fails on a less
operationalist answer (favored by one of us \cite{healey}).
On this latter, pragmatist approach, one is licensed to infer from a system's being assigned
an eigenstate of an observable to its having the corresponding eigenvalue
only in so far as the joint state of system and environment has decohered
with eigenstates of the observable corresponding to a ``pointer basis''. In the situation described by Lalo\"{e} the surprising thing is also that apparently
a microscopic interaction (the measurement of a relatively small number of
particles) has a macroscopic outcome: the appearance of a large
angular momentum. The same paradox of a very small quantity flipping a
macroscopic one appears in the analysis of the Josephson effect \cite{leggettsols}.

Similarly in the case of the Bose-Einstein double condensate, after a small number of transversal measurements, we end up in a state which approximates an eigenstate
of the angular momentum along a direction in the $xOy$ plane with a large
eigenvalue. Again, we tend to imagine this state and the corresponding
angular momentum existing. Just as in the Bloch sphere representation, where
the eigenvectors of the spin can be mathematically identified with arrows
pointing in some direction, we tend to believe that these arrows represent
something real. But the arrows correspond to quantum state assignments, not
assignments of values to observables \emph{in} those states. Even for an
assignment of an angular momentum eigenstate to a system, it is only at the end of a measurement that we can say that the system
\emph{has} an angular momentum component in the direction of the arrow equal
to the corresponding eigenvalue.
\textit{There is no angular momentum before it was measured.}

One can object to all the above by saying; but
isn't it the case that the total spin of a ferromagnet is ``real''? Isn't it
the case that the existence of a ``real'' angular momentum is established by
experiments such as Einstein - de Haas? To this we can only answer:
all experiments in physics end with a measurement.  The lesson we learn from quantum physics is that it is real because it is
has a measured value. It is not the case that something has a measured value because it is
real. In many practical situations in physics we do not need to distinguish
between the two: but sometimes we are forced to admit that they are
different.

\section{Acknowledgements}

 The research for this work was initially funded by the Templeton Research Fellows Program ``Philosophers -
 Physicists Cooperation Project on the Nature of Quantum Reality'' at the Austrian Academy of Science's Institute for Quantum Optics and Quantum Information, Vienna, which allowed the authors to spend the summer of 2009 at IQOQI Vienna. Special thanks go to our
hosts in Vienna, Prof. A. Zeilinger and Prof. M. Aspelmeyer, who have
made this visit possible, and additionally to the scientists in the institute for many enlightening discussions.  Also, GSP acknowledges support from the Academy of Finland (Acad. Res. Fellowship 00857, and projects 129896, 118122, 135135, and 141559). The contribution by RH to this material is partially based upon work supported by the National Science Foundation under Grant No. SES-0848022.


\begin{thebibliography}{9}

\bibitem{laloe} F. Lalo\"e: Bose-Einstein condensates and EPR quantum
non-locality. In: T.M. Nieuwenhuizen \textit{et. al.} (eds.) Beyond the
Quantum, pp. 35-52. World Scientific, Singapore (2007); arXiv:0704.0386.

\bibitem{ketterle} Andrews, M.R., \textit{et. al.}: Observation of
interference between two Bose condensates, Science \textbf{275}, 637-641
(1997).

\bibitem{theory} J. Javanainen and S. M. Yoo: Quantum phase of a
Bose-Einstein condensate with an arbitrary number of atoms. Physical Review
Letters \textbf{76}, 161-4 (1996); Castin, Y. and Dalibard, J.: Relative
phase of two Bose-Einstein condensates. Physical Review A \textbf{55},
4330-7 (1997); Lalo\"e, F.: The hidden phase of Fock states; quantum non-local effects. The
European Physics Journal D \textbf{33}, 87-97 (2005); Mullin, W.J., Krotkov,
R. and Lalo\"e, F.: Evolution of additional (hidden) quantum variables in the
interference of Bose-Einstein condensates. Physical Review A \textbf{74},
023610:1-11 (2006); Mullin, W.J., Krotkov, R. and Lalo\"e, F.: The origin of the phase in the interference of Bose-Einstein
condensates. American Journal of Physics {\bf 74}, 880-87 (2006); Lalo\"e, F. and Mullin, W.J.: Einstein-Podolsky-Rosen argument and Bell
inequalities for Bose-Einstein spin condensates. Physical Review A \textbf{77%
}, 022108:1- 17 (2008); Paraoanu, G.S.: Localization of the relative phase
via measurements. Journal of Low Temperature Physics \textbf{153}, 285-93
(2008).

\bibitem{EPR} A. Einstein, B. Podolsky, and N. Rosen: Can quantum-mechanical description of  reality be considered complete? Phys. Rev. {\bf 47}, 777-780 (1935).

\bibitem{Bohm} D. Bohm: (1951) Quantum Theory. Prentice-Hall, New York.

\bibitem{GHZ} D. M. Greenberger, M. A. Horne, and A. Zeilinger, (1989) in
Bell's Theorem, Quantum Theory, and Conceptions of the Universe, ed. M.
Kafatos, 73-76. Kluwer, Dordrecht.




\bibitem{healey} R. Healey: Quantum Theory: a Pragmatist Approach, forthcoming in the British Journal for the Philosophy of Science,
arXiv:1008.3896.




\bibitem{peresbohr}
Peres, A.: Quantum Theory: Concepts and Methods. Dordrecht: Kluwer Academic (1993).

\bibitem{leggettsols} Leggett, A.J. and Sols, F.: On the concept of
spontaneously broken gauge symmetry in condensed matter physics. Foundations
of Physics \textbf{21}, 353-364 (1991).











\end{thebibliography}
\end{document}